\documentclass[journal=jctcce,manuscript=article,layout=twocolumn]{achemso}
\setkeys{acs}{etalmode=truncate}
\setkeys{acs}{maxauthors=3}
\setkeys{acs}{articletitle=true}
\usepackage[version=3]{mhchem} 
\usepackage[T1]{fontenc}       
\usepackage{amsfonts,amsmath,amsbsy,color,enumitem}
\usepackage[]{graphics,graphicx}
\usepackage{subcaption}
\usepackage{bm}

\usepackage[colorinlistoftodos]{todonotes}
\usepackage[colorlinks,citecolor={blue}]{hyperref}
\usepackage[capitalise]{cleveref}

\newcommand{\D}[1] {\mathrm{D}_{\bf #1}}

\newcommand{\braket}[3] {{\langle #1 | #2 | #3 \rangle}}
\newcommand{\brket}[2] {{\langle #1 | #2 \rangle}}

\newcommand{\ket}[1] {{| #1 \rangle}}

\author{Verena~A.~Neufeld}
\email{verena.a.neufeld@gmail.com}
\author{Alex~J.~W.~Thom}
\email{ajwt3@cam.ac.uk}
\affiliation{Department of Chemistry, Lensfield Road, Cambridge CB2 1EW, United Kingdom}
\date{\today}

\title{Accelerating Stochastic Quantum Chemistry}

\keywords{Quantum Chemistry, Computational Optimization, Monte Carlo, Coupled Cluster, Full Configuration Interaction}

\begin{document}

\begin{tocentry}
\includegraphics[width=4.7cm,keepaspectratio]{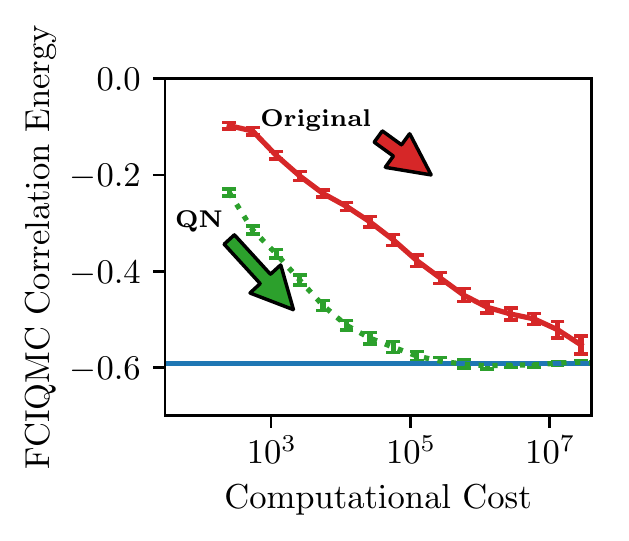}
\end{tocentry}

\begin{abstract}
The convergence of full configuration interaction quantum Monte Carlo (FCIQMC) is accelerated using a quasi-Newton propagation (QN) which can also be applied to coupled cluster Monte Carlo (CCMC). The computational scaling of this optimised propagation is $\mathcal{O}(1)$, keeping the additional computational cost to a bare minimum. Its effects are investigated deterministically and stochastically on a model system, the uniform electron gas, with Hilbert space size up to $10^{40}$ and shown to accelerate convergence of the instantaneous projected energy by over an order of magnitude in the FCIQMC test case. Its capabilities are then demonstrated with FCIQMC on an archetypical quantum chemistry problem, the chromium dimer, in an all-electron basis set with Hilbert space size of about $10^{22}$ yielding highly accurate FCI energies.
\end{abstract}

\section{Introduction}
\label{sec:intro}
Accurate energies of electronic systems are not only crucial as benchmarks and verification for other, often computationally cheaper, methods (see e.g. Refs. \citenum{Bartlett2007, Lynch2003a,Al-Hamdani2014,Veit2019}) but are also needed
when knowledge of those energies to a very high accuracy and precision is desired, for example when determining the low-energy crystal/molecular structures (e.g. \citenum{Yang2014,Mostaani2015,Gruber2018}) or vibrational spectra (e.g. Refs. \citenum{El-Azhary2003,Christiansen2004,Yuwono2019}).
Wavefunction based quantum chemistry methods, such as (full) configuration interaction, (F)CI,\cite{Shavitt1977,Cremer2013} or
coupled cluster theory, CC,\cite{Coester1960,Cizek1966,Cizek1971,Bartlett2007} can give accurate, if not exact, energies. This accuracy is systematically improvable
unlike density functional theory\cite{Kohn1965a,Hohenberg1964} and they do not require \emph{ab initio} knowledge of
the wavefunction as diffusion Monte Carlo\cite{Foulkes2001} does. Coupled cluster at the level of singles doubles
and perturbative triples, CCSD(T),\cite{Raghavachari1989} has been shown to be able to give chemical accuracy, 1 kcal/mol, in some systems and is said
to be the ``gold standard''\cite{Bartlett2007}.
\par In the last decade, Booth et al.\cite{Booth2009} and Thom\cite{Thom2010} introduced highly parallelisable\cite{Booth2014,Spencer2018a} stochastic
versions of FCI, FCIQMC, and CC, CCMC, respectively, reducing memory costs and thus enabling calculations at larger basis
sets.
Not just molecules have been tackled with FCIQMC and CCMC (e.g. see Refs. \citenum{Booth2011, Cleland2012, Daday2012, Booth2014, Holmes2016a,Sharma2014,Veis2018a,Samanta2018,Thom2010,Spencer2016,Spencer2018a}), but also the uniform electron gas\cite{MartinUEGChapter,Shepherd2012a,Shepherd2013, Spencer2016, Shepherd2016a, Neufeld2017b,Luo2018a} or realistic solids\cite{Booth2013,Spencer2019c} for example. Besides accurate ground state energies, FCIQMC has given excited state energies as well\cite{Booth2012,Ten-no2013,Humeniuk2014,Blunt2015,Blunt2017}. Some of its great benefits were demonstrated when FCIQMC reached systems of extremely large Hilbert space sizes using
the initiator approximation\cite{Booth2014,Cleland2010,Shepherd2012a}, and other improvements of the algorithms
have further increased the sampling efficiency of FCIQMC and/or CCMC 
\cite{Petruzielo2012,Blunt2015b,Holmes2016a,Neufeld2018a,Spencer2018a,Scott2017b}, with a recent diagrammatic CCMC version increasing efficiencies by moving CCMC closer to conventional coupled cluster\cite{Scott2019a}.
\par An irksome problem faced by FCIQMC and CCMC has not yet been largely discussed in the literature: the computational
effort to reach equilibration can be very significant and thus it can take a prohibitive amount of time
before the energy can be even roughly estimated.
Here, we introduce a modification to the algorithms: the quasi-Newton method which is commonly used in conventional deterministic coupled
cluster\cite{Helgaker2000}. This is closer to using the quadratically convergent Newton--Raphson optimisation to reach
equilibration instead of the linearly convergent steepest descent method. The Hessian required may be approximated by using inexpensive Fock expectation value sums.
Since this method has been developed and implemented\cite{Spencer2019c}\footnote{See \url{https://github.com/hande-qmc/hande} for code.} for CCMC and FCIQMC,
Blunt et al.\cite{Blunt2019b} have also introduced an alternative Jacobi pre-conditioned propagation\cite{Davidson1975}. A comparison is made to their
method which is computationally more expensive than the approach presented here. Note that other propagator improvements, which are not discussed here, exist, including the use of
Chebyshev expansion\cite{Zhang2016} and techniques used in the machine learning community which have also been applied to Quantum Monte Carlo methods to accelerate convergence\cite{Schwarz2017,Sabzevari2018a,Otis2019a}.
Deustua et al.\cite{Deustua2017,Deustua2018} have used FCIQMC and CCMC to estimate deterministic amplitudes/coefficients and managed to converge to highly accurate energies quickly doing so, see for
example the CAD-FCIQMC method\cite{Deustua2018}. This approach is orthogonal to the convergence acceleration shown here, in fact they can be most likely
employed simultaneously to improve convergence.
\par First, we will describe the quasi-Newton propagation, followed by analysing its convergence behaviour in both
\textit{deterministic} and \textit{stochastic} propagations and comparing it to the original, Jacobi, and full Newton propagations.
Finally, the quasi-Newton propagation is applied to the chromium dimer in the full Ahlrichs' SV basis\cite{Schafer1992} demonstrating its capabilities for accurate calculations of large quantum chemical systems.

\section{Theory}
The quasi-Newton propagation formalism is derived by treating FCIQMC as an
optimisation problem. The derivation is similar to a derivation by Davidson\cite{Davidson1975}. The conclusion also holds for CCMC. Past
literature\cite{Booth2009,Cleland2010,Thom2010,Spencer2016,Spencer2018a,Spencer2019c} contains detailed
descriptions of the CCMC and FCIQMC algorithms.
\par In FCIQMC, the lowest eigenvalue of the Hamiltonian is found along with an approximation of its eigenvector
which is the ground state wavefunction. Working in Slater determinant space, the ground state
wavefunction can be written as $\Psi_0 = \sum_{\mathbf{i}} c_\mathbf{i} \ket{\D{i}}$ where
$\mathbf{i}$ is a unique label for (known) determinant $\ket{\D{i}}$ and its corresponding (unknown)
coefficient $c_\mathbf{i}$ that is determined using FCIQMC. The constraint is the
normalisation of the wavefunction, $\brket{\Psi}{\Psi} = N$ for some constant $N$. A Lagrangian $\mathcal{L}$ with Lagrange multiplier $E$ can therefore be written as
\begin{equation}
\mathcal{L} =  \braket{\Psi}{\hat{H}}{\Psi}- E(\brket{\Psi}{\Psi} - N).
\end{equation}
Differentiating gives the gradient
\begin{equation}
g_\mathbf{i} = \frac{\partial\mathcal{L}}{\partial c^*_\mathbf{i}} \propto  \braket{\D{i}}{\hat{H} - E}{\Psi}.
\end{equation}
Setting $\bm{g}$$=$$\bm{0}$ gives the converged (F)CI equations $\braket{\D{i}}{\hat{H} - E}{\Psi}=0$ for all $\mathbf{i}$. In
the original FCIQMC formalism\cite{Booth2009}, $g_\mathbf{i} = \braket{\D{i}}{\hat{H} - E}{\Psi}$ is used to propagate from the initial guess to the ground state wavefunction in imaginary time, $\tau$, with an update equation equivalent to steepest descent,
\begin{equation} 
\bm{c}(\tau + \delta \tau) = \bm{c}(\tau) - \delta \tau \bm{g}(\tau)
\end{equation}
using time step $\delta \tau$. The optimised wavefunction is $\Psi_0$ with energy $E$.
\par Steepest gradient descent approaches the solution linearly and is therefore inefficient. The quadratically convergent Newton--Raphson method propagates the coefficients towards $\bm{g}$$=$$\bm{0}$ by
\begin{equation}
\bm{c}(\tau + \delta \tau) = \bm{c}(\tau) - \delta \tau \bm{\mathrm{\tilde{H}}^{-1}}\bm{g}(\tau)
\end{equation}
where the time step $\delta \tau$ was retained for extra flexibility. The elements of the Hessian $\bm{\mathrm{\tilde{H}}}$ are given by
\begin{equation}
\tilde{H}_{\mathbf{i}\mathbf{j}} = \frac{\partial g_\mathbf{i}}{\partial c_\mathbf{j}} \propto \braket{\D{i}}{\hat{H} - E}{\D{j}}.
\end{equation}
Since inverting $\bm{\mathrm{\tilde{H}}}$ is highly expensive, approximations to $\bm{\mathrm{\tilde{H}}^{-1}}$ are necessary. It may be assumed that the off-diagonal elements in
$\bm{\mathrm{\tilde{H}}}$ are not very significant compared to the diagonal elements and so can be set to
zero, leaving an easily invertible diagonal matrix, provided no diagonal elements are zero. Davidson\cite{Davidson1993} has noted the connection of pre-conditioning to the Newton--Raphson algorithm; while derived differently, this is equivalent to
the Jacobi pre-conditioned propagation used by Blunt et al.\cite{Blunt2019b}.
\par Here, the example of coupled cluster theory is followed\cite{Helgaker2000} where Fock expectation values for orbitals $i$, $\braket{i}{\hat{F}}{i}$, are used in an approximation to the
diagonal Hamiltonian elements and off-diagonal elements are ignored. The diagonal elements of $\bm{\mathrm{\tilde{H}}}$,
$\propto \braket{\D{j}}{\hat{H} - E_\mathrm{HF} - E_\mathrm{proj.}}{\D{j}}$, are approximated by the sum of Fock expectation values of occupied orbitals in $\D{j}$ minus the sum of Fock expectation values  of occupied orbitals in the reference,
\begin{equation}
\begin{split}
&\braket{\D{j}}{\hat{H} - E}{\D{j}} \approx\\ &\sum_{m\ \mathrm{in}\ \bm{j}} \braket{m}{\hat{F}}{m} - \sum_{m^\prime\  \mathrm{in}\ \bm{0}} \braket{m^\prime}{\hat{F}}{m^\prime}.
\end{split}
\label{eq:focksum}
\end{equation}
Note that the
computational cost of Blunt's Jacobi pre-conditioned propagation\cite{Blunt2019b} is at least $\mathcal{O}(N_\mathrm{el.})$\footnote{To approximate
$\bm{\mathrm{\tilde{H}}}$, we need to evaluate $\braket{\D{j}}{\hat{H} - E}{\D{j}}$ as part of the
death step for any type of propagation, so there is no extra cost in the death step. For the spawn
step, $\braket{\D{i}}{\hat{H} - E}{\D{i}}$ is needed. Since $\D{i}$ and $\D{j}$ differ by at most a
double excitation, $\braket{\D{j}}{\hat{H} - E}{\D{j}}$ can be used as a starting point and the difference can be calculated. This is an $\mathcal{O}(N_\mathrm{el.})$ operation (Personal communication with Dr. Nick Blunt).} whereas the computational cost due to the quasi-Newton propagation is $\mathcal{O}(1)$.

\section{Deterministic Propagation}
To test this approximation, the different propagation techniques were first deterministically tested on a small model system where the true eigenvalues and eigenvectors were known, and stochastic noise, reaching the level of a 
sufficient number of particles and other challenges in stochastic propagations, could be ignored so
the focus was solely on how many iterations were needed to converge.
\par The model system studied was the three-dimensional uniform electron gas (UEG) with two electrons of opposite spin in
1850 spinorbitals which has a Hilbert space size of 925. There, the Fock value for spinorbital $m$ is given by\cite{ShepherdThesis2013}
\begin{equation}
\begin{split}
&\braket{m}{\hat{F}}{m}=\\ &\frac{1}{2}|\bm{k}|^2 - \sum_{\substack{n \ \mathrm{in}\ \bm{0}\\ m \neq n\\
\mathrm{same}\ \mathrm{spin}}} \left\langle nm\left|\frac{1}{\bm{r_{12}}}\right|mn\right\rangle \left(+ \frac{1}{2}V_\mathrm{Mad.}\right)
\end{split}
\end{equation}
where the last term in round brackets including the Madelung constant per electron $V_\mathrm{Mad.}$ is added to spinorbitals $m$ occupied in the reference only. $V_\mathrm{Mad.} \approx -2.837297\times(\frac{3}{4\pi r_s^3N_\mathrm{el.}})^{1/3}$ as determined by Schoof et al.\cite{Schoof2015,Fraser1996} with Wigner-Seitz radius $r_s$.
Using the HANDE QMC code\cite{Spencer2019c}, an FCI calculation was performed and Hamiltonian matrix elements were calculated. 
\begin{figure}
\centering
	\begin{subfigure}[h]{8.5cm}
	\includegraphics[width=1.0\linewidth,keepaspectratio]{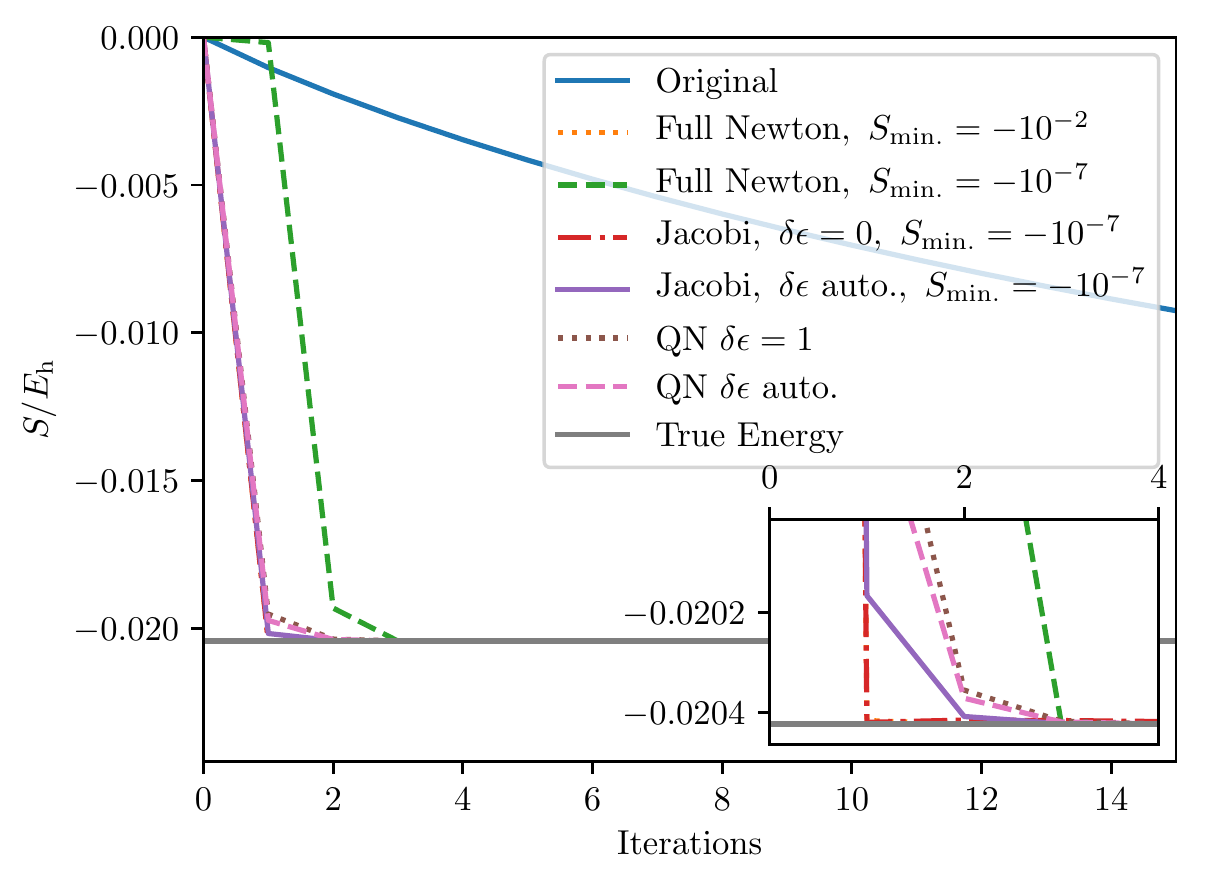}
	\caption{Propagation for $r_s = 0.5\mathrm{a}_\mathrm{0}$.}
	\label{fig:rs0.5}
	\end{subfigure}
	\newline
	\begin{subfigure}[h]{8.5cm}
	\includegraphics[width=1.0\linewidth,keepaspectratio]{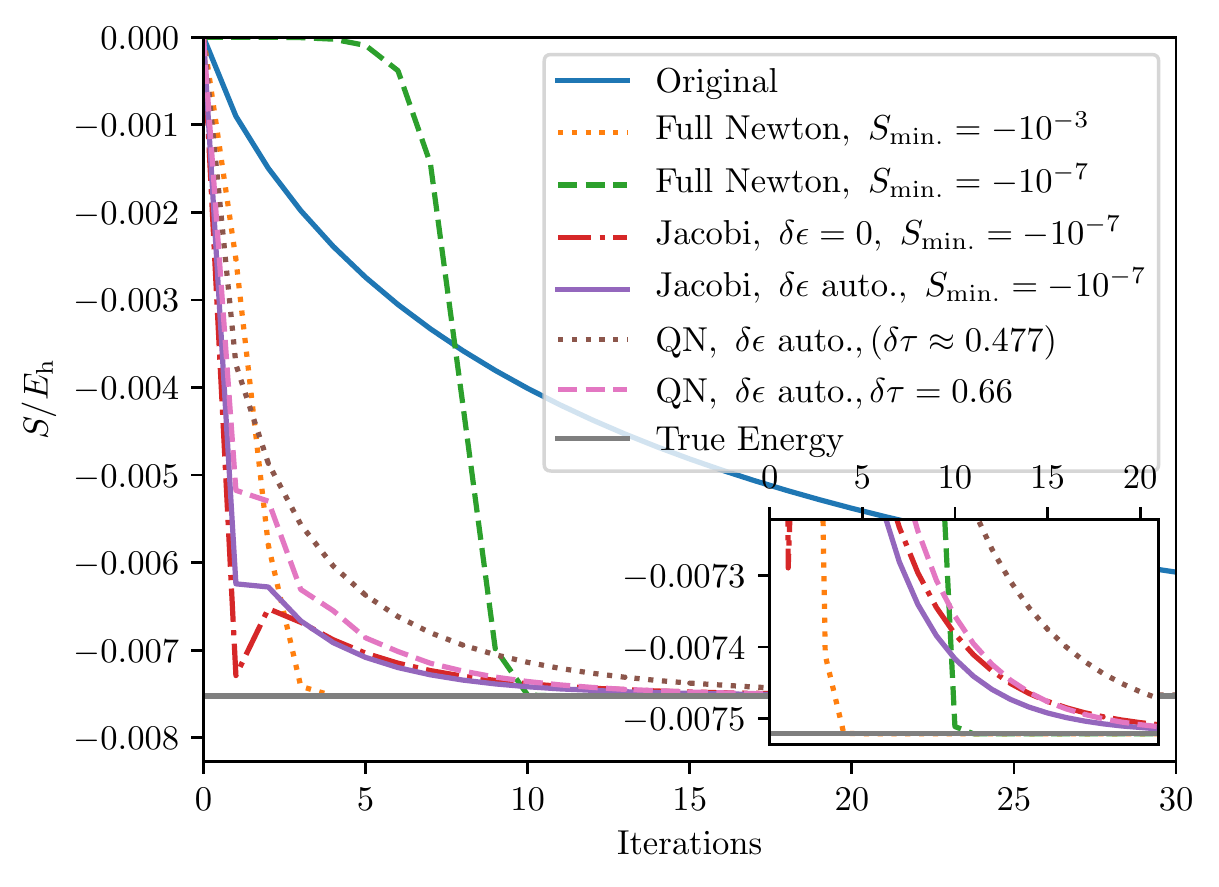}
	\caption{Propagation for $r_s = 20\mathrm{a}_\mathrm{0}$.}
	\label{fig:rs20}
	\end{subfigure}
	\newline
	\caption{\small{Deterministic propagation of the 3D UEG with two electrons of opposite spins in 1850 spinorbitals with
	original, full Newton, Jacobi pre-conditioned and quasi-Newton propagation for $r_s = 0.5\mathrm{a}_\mathrm{0}$ (a)
	and $r_s = 20\mathrm{a}_\mathrm{0}$ (b). Jacobi $\delta \epsilon$ auto. sets the first diagonal element of the approximated Hessian to its second element whereas $\delta \epsilon = 0$ leaves the diagonal untouched. The shift is set to the projected energy at each iteration and the time step (except for the quasi-Newton run with $\delta \tau = 0.66$) for each propagation is the reciprocal of the highest eigenvalue of the propagation matrix $\bm{\mathrm{A}}^{-1}\bm{\mathrm{\tilde{H}}}$. In case of full Newton and Jacobi propagations,  $S=S_\mathrm{min.}$ if $|S| < |S_\mathrm{min.}|$. The full Newton curve with $S_\mathrm{min.}=-10^{-2}$ and the Jacobi one with $\delta\epsilon=0$ cannot be distinguished at this scale.}}
	\label{fig:detprop}
\end{figure}
The initial guess for the wavefunction was a vector with 1 at the $\D{0}$ position and 0 otherwise.
This corresponds to a standard FCIQMC calculation with initially one Monte Carlo particle at the reference
determinant. The shift $S$ was set to the projected energy at every iteration. The time step was set
to the reciprocal of the highest eigenvalue of $\bm{\mathrm{A}}^{-1}\bm{\mathrm{\tilde{H}}}$ where in the original propagation 
$\bm{\mathrm{A}}$ is the identity and in the other propagations it is the Hessian $\bm{\mathrm{\tilde{H}}}$
or an approximation thereof. This was inspired by the fact that the highest
allowed time step in FCIQMC is twice the reciprocal of the highest eigenvalue of
$\bm{\mathrm{\tilde{H}}}$\cite{Booth2009,Trivedi1990} although this might not apply to all propagations exactly. The full Newton propagation used a Hessian
with elements $\braket{\D{i}}{\hat{H} - 0.99S}{\D{j}}$ with a factor of 0.99 since its inverse would
otherwise tend to be singular as $S\rightarrow E_\mathrm{corr.}$. In the first iteration, where the
shift and projected energy are zero, the first diagonal element is set to a small number such as $10^{-2}$, $10^{-3}$ or $10^{-7}$
(see figure \ref{fig:detprop}),
in the case of the full Newton and Jacobi propagations. If \textit{auto.} mode is chosen when using the quasi-Newton propagation, the first diagonal
element of the approximated Hessian would be zero, so it is set to the second diagonal element. When using the Jacobi propagation, $E$ in the propagation is set to the shift $S$ and a threshold $\delta \epsilon$ is applied or the first element set to the second (auto. $\delta \epsilon$).
Figure \ref{fig:detprop} shows the propagation for $r_s = 0.5\mathrm{a}_\mathrm{0}$ and $r_s =
20\mathrm{a}_\mathrm{0}$. For quasi-Newton, two time steps are shown; one found as described above ($\approx$0.477), and the other being 0.66 which is higher.
\par Clearly, in terms of convergence with respect to iterations, the original propagation is
outperformed by the others which perform similarly to each other. As demonstrated by the full Newton propagation the initial guess for $S=S_\mathrm{min.}$ can obviously affect convergence. The higher time step used for quasi-Newton performs slightly better than the automatically found time step but it is still similar in behaviour. The more correlated the UEG
system gets, the higher $r_s$, the smaller the range in Fock eigenvalues so the more similar the
original propagation is to the quasi-Newton propagation. 

\section{Stochastic Propagation}
Next, the quasi-Newton propagation is compared with the original propagation in FCIQMC.
The quasi-Newton propagation can be straightforwardly implemented into FCIQMC as the only changes are
in the \textit{spawn} and \textit{death} steps. In the case of the \textit{spawn} step,
the probability that a spawn is accepted
is divided by $\Delta_{\bf{i}}$ where

\begin{equation}
\Delta_{\bf{i}}=
\begin{cases}
\Delta_{\bf{i}}^\prime & \mathrm{if}\ \Delta_{\bf{i}}^\prime \geq \delta \epsilon \\
\Delta_\mathrm{v} & \mathrm{otherwise}
\end{cases}
\end{equation}
with
\begin{equation} 
\Delta_{\bf{i}}^\prime=\sum_{m\ \mathrm{in}\ \bm{j}} \braket{m}{\hat{F}}{m} -
\sum_{m^\prime\  \mathrm{in}\ \bm{0}} \braket{m^\prime}{\hat{F}}{m^\prime}.
\end{equation}
$\delta \epsilon$ is a threshold and $\Delta_\mathrm{v}$ an alternative value chosen which could be set to 1 (see later part on the chromium dimer) or, as in this stochastic UEG study here, to $\delta \epsilon$. Similarly to the deterministic investigation,
$\delta \epsilon$ can be chosen to be the difference between the sum of Fock energies of the reference and
first excited determinant to maximise the time step possible. In the original \textit{death} step, the
\textit{death} probability of a particle on determinant $\ket{\D{i}}$ is written as\cite{Booth2009}
\begin{equation}
p_\mathrm{death}(\ket{\D{i}}) \propto \delta \tau \braket{\D{i}}{\hat{H} - S}{\D{i}}.
\end{equation}
If a quasi-Newton modification were also performed to the \textit{death} step, the resulting \textit{death} probability
would be $\frac{p_\mathrm{death}(\ket{\D{i}})}{f_{\bf{i}}}$. We consider the hypothetical case where the estimate of the wavefunction is a multiple of the true wavefunction,
but $S$ is not equal to the true energy the wavefunction would stay at the true solution as all determinants are affected equally by the error in $S$ in the
\textit{death} step. However, in the case of quasi-Newton, due to the determinant dependence of $\Delta_{\bf{i}}$, the
estimate of the wavefunction would move away from the true solution. A modified \textit{death} step (inspired by the coupled cluster Monte Carlo modification of Franklin et al. \cite{Franklin2016}) is
\begin{equation}
\begin{split}
&p_\mathrm{death}(\ket{\D{i}}) \propto \\ &\delta \tau \left(\frac{\braket{\D{i}}{\hat{H} - E_\mathrm{proj.}}{\D{i}}}{\Delta_{\bf{i}}} + \rho (E_\mathrm{proj.} - S)\right),
\end{split}
\label{eq:pdeath}
\end{equation}
with the projected energy $E_\mathrm{proj.}$ and $\rho$ as a constant population control factor to
add an extra degree of freedom. We have assumed that $E_\mathrm{HF}$ has already been subtracted of the Hamiltonian matrix diagonal. At the true solution,
$E_\mathrm{proj.}$ takes the correct value so the net effect of the first term in equation \ref{eq:pdeath} when applied to the whole population is zero, and the latter term merely scales the whole population, so the wavefunction remains at the true solution.
\par
\begin{figure}[t!]
\centering
	\includegraphics[width=8.5cm,keepaspectratio]{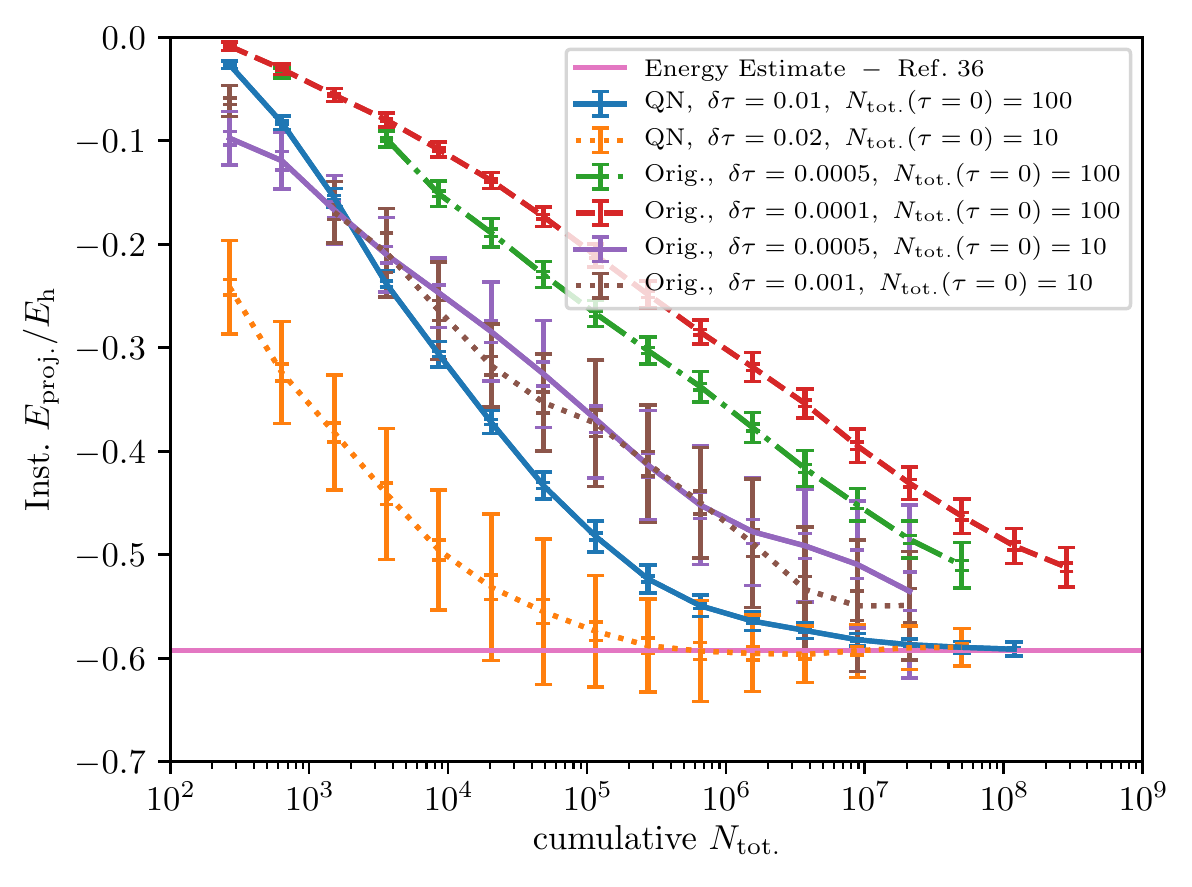}
	\caption{\small{Convergence of the instantaneous projected energy as a function of cumulative number of
	Monte Carlo particles $N_\mathrm{tot.}$ as cost measure for the 3D UEG with 1850 spinorbitals, 14 electrons, and $r_s
	= 0.5\mathrm{a}_\mathrm{0}$ in FCIQMC. The instantaneous projected energy was binned with respect
	to the cumulative particle number and the bin means calculated. Each calculations was done at least 20 times
	in independent runs and the means of those bin means are shown with their standard deviations
	and standard errors as outer and inner error bars respectively, placed at the number of cumulative
	particles that is at the middle of the bin. Not every data point was determined by the same number
	of independent bins. Calculations here were stopped pre-maturely and could have continued, so the end of a curve does not imply that the
	memory was full.
	The estimate for the true energy using (initiator) FCIQMC was taken from Ref.
	\citenum{Neufeld2017b}. Twice its error is shown in the line spread but it is too small to be
	visible. $N_\mathrm{tot.}(\tau = 0)$ is the initial population. $\rho = 1.0$, $\delta
	\epsilon \approx 11.8 E_\mathrm{h}$, $\Delta_\mathrm{v} = \delta \epsilon$ and the shift was not
	varied.}}
	\label{fig:instprojeparticlesconverging}
\end{figure}

 Using the spin non-polarised three dimensional (3D) UEG again, this time with 1850 spinorbitals, 14 electrons, and $r_s =
0.5\mathrm{a}_\mathrm{0}$, the stochastic propagations using FCIQMC with quasi-Newton and the
original propagation were compared. The instantaneous projected energies were binned with respect to
the cumulative number of particles, $N_\mathrm{tot.}$, to reach those instantaneous projected energies and the mean in
each bin for each calculation run calculated. The same calculation was then run at least 20 times with
different random number generator seeds. The means of these independent bin means are shown in
figure \ref{fig:instprojeparticlesconverging} with their standard deviations and standard errors
across the different runs as error bars. Empty bins did not contribute to the mean or its errors.
The bin positions are the same for all calculations. Note that not all calculations ran for the same number of iterations, some ended early.
The cumulative number of particles $N_\mathrm{tot.}$ is a measure of the cost of
the calculation that is more implementation-- and platform--independent than the compute time for
example, as an iteration in the FCIQMC algorithm scales approximately linearly in the number of
particles at that time step\footnote{Each particle does one spawn attempt here.}. A pre-calculated $\mathcal{O} (1)$ version of a \textit{uniform}
\textit{Power--Pitzer} excitation generator adapted to the UEG was
used\cite{Neufeld2018a,Smartunpub}. Floating-point amplitudes\cite{Petruzielo2012,Overy2014} were employed with a spawn cutoff of 0.01.
Figure \ref{fig:instprojeparticlesconverging} shows that the instantaneous projected energy
converges significantly faster when using the quasi-Newton propagation. The time steps for the quasi-Newton
propagation are 10--40 times greater than time steps of the original propagations shown. $\delta
\epsilon \approx 11.8 E_\mathrm{h}$, the Fock value difference between the ground and first excited
determinant of the same symmetry. As expected, using a lower initial population decreases the
initial cost of converging to a certain energy but increases the noise. Population control has not been applied here, we have just focussed on convergence, not 
evaluating the final energy.
\par To test how the system size affects the performance of quasi-Newton compared to the original propagation, i.e. whether quasi-Newton can be (even) more 
beneficial in larger systems, the convergence for the same 3D uniform electron gas but with 11150 spinorbitals was investigated and compared to the 1850 
spinorbitals case shown in figure \ref{fig:instprojeparticlesconverging}. This is a system with Hilbert space size of about $10^{40}$. The memory capacity of the
spawn array was fixed and calculations were allowed to increase their population until this array was full and the calculation was then stopped (and the last
iteration printed disregarded).
\begin{figure}[t!]
\centering
	\includegraphics[width=8.5cm,keepaspectratio]{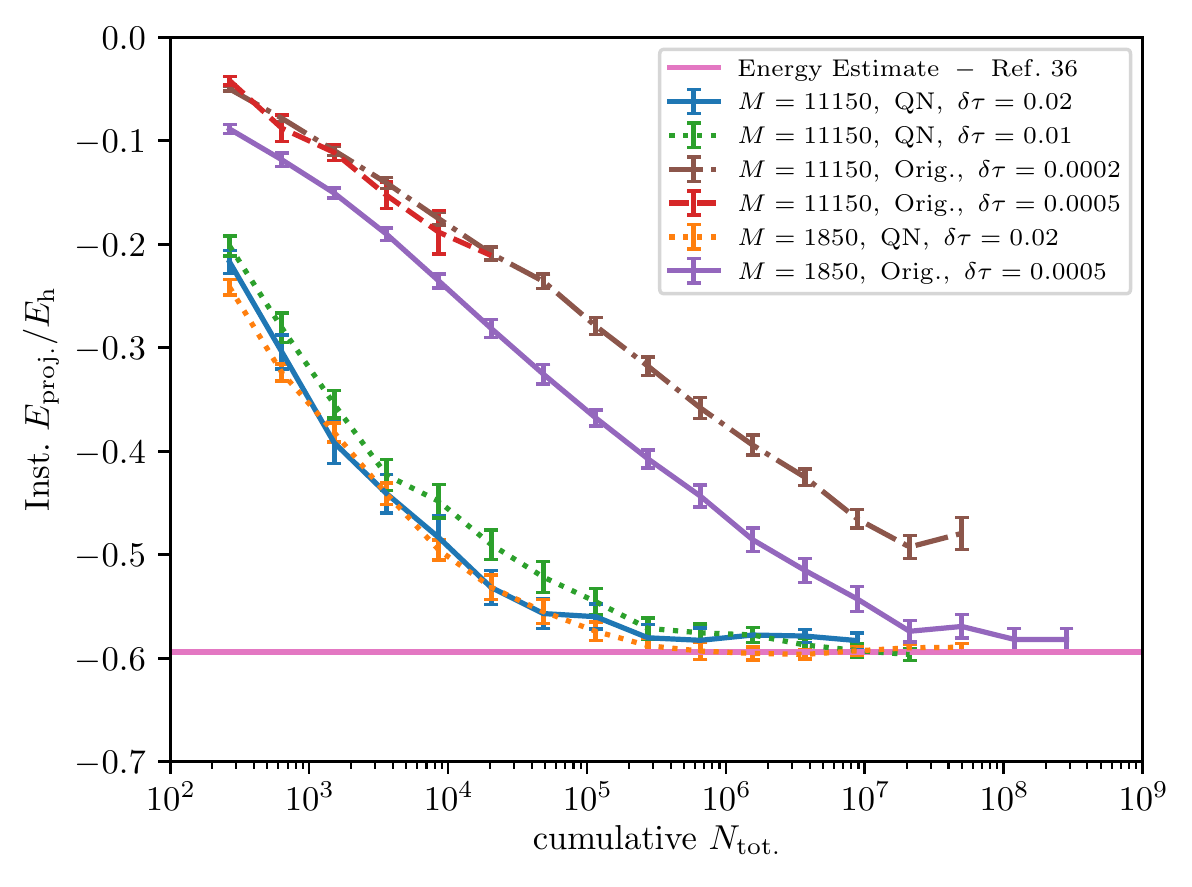}
	\caption{\small{Convergence of the instantaneous projected energy as a function of cumulative number of
	Monte Carlo particles $N_\mathrm{tot.}$ as cost measure for the 3D UEG with $M=$ 11150 spinorbitals, 14 electrons, and $r_s
	= 0.5\mathrm{a}_\mathrm{0}$ in FCIQMC. The instantaneous projected energy was binned with respect
	to the cumulative particle number and the bin means calculated. Each calculations was done at least 20 times
	in independent runs and the means of those bin means are shown with their standard errors as error bars respectively (unlike figure \ref{fig:instprojeparticlesconverging}, standard deviations are not shown), placed at the number of cumulative
	particles that is at the middle of the bin. Not every data point was determined by the same number
	of independent bins. Unlike in figure \ref{fig:instprojeparticlesconverging}, calculations here were only stopped pre-maturely if the array containing the 
	spawned particles was full and the last printed iteration ignored, except for the quasi-Newton propagation shown for comparison run at 1850 spinorbitals. The calculations for the original propagation at 1850 spinorbitals were rerun.
	Two estimates for the true energy using (initiator) FCIQMC, taken from Ref.
	\citenum{Neufeld2017b}, are shown; the energy at 4218 spinorbitals and the complete basis set extrapolated limit. However, they cannot be distinguished 
	within the thickness of the horizontal line. Twice their error is shown respectively in the line spread but it is too small to be
	visible. $N_\mathrm{tot.}(\tau = 0)$, the initial population, was 10. $\rho = 1.0$, $\delta
	\epsilon \approx 11.8 E_\mathrm{h}$, $\Delta_\mathrm{v} = \delta \epsilon$ and the shift was not
	varied.}}
	\label{fig:instprojeparticlesconverging60.5}
\end{figure}
For clarity, only the standard errors are shown this time as errorbars. One of the fastest converging curves each for each propagation run at $M = 1850$ is 
shown as well for direct comparison. In terms of convergence, the quasi-Newton propagation does not appear to be strongly affected by the increase in system 
size. However, the original propagation converges slower, not being able to contain the spawns in the given --- fixed --- spawn array. The original propagation
therefore either requires more memory or has to lower the time step which in turn, as shown by figure \ref{fig:instprojeparticlesconverging}, decreases
the rate of convergence. Considering that the orbital Fock value increases for added sets of spinorbitals, the stabilising behaviour of quasi-Newton was to be 
expected by considering that the spawn probability effectively is divided by the difference in the sum of Fock energies to the reference sum.
Provided there is an adequate range in Fock energies, quasi-Newton therefore enables faster convergence rates --- especially in larger systems. 

\section{Application to \ce{Cr2}}
Finally, the quasi-Newton propagation was tested on an archetypical quantum chemistry problem, the chromium dimer, at a bond length of 1.5\AA. The basis set considered is Ahlrichs' SV\cite{Schafer1992} where first a CAS
of 24 electrons correlated in 30 spatial orbitals was applied and then the full system was studied with initiator FCIQMC. The Hartree--Fock orbitals and their integrals were evaluated with the Psi4 code\cite{Turney2012,Parrish2017}. The weighted
heat-bath excitation generator\cite{Holmes2016a} (adapted\cite{Neufeld2018a}) has been used. Again, floating-point amplitudes\cite{Petruzielo2012,Overy2014} were employed with a spawn cutoff of 0.01. Booth et
al.\cite{Booth2014} have previously applied FCIQMC to the chromium dimer with a CAS and DMRG results exist for
both smaller CAS\cite{Kurashige2009,Sharma2012,Olivares-Amaya2015} and full\cite{Olivares-Amaya2015} system, also in Ahlrich's SV basis\cite{Schafer1992}\footnote{Refs.\citenum{Sharma2012} and \citenum{Booth2014} state that they have used Ahlrich's SV(P) or SVP basis set. In summary, given that their results agree very well with ours, we conclude that we most likely used the same, SV basis set, details given here. The basis we used (Ahlrich's SV basis set) can be found at EMSL  Basis Set Exchange Library, \url{https://bse.pnl.gov/bse/portal} [accessed 22.05.2019], under ``Ahlrichs VDZ'' and selecting ``Cr'' as the element. It has \{63311/53/41\} functions\cite{Schafer1992}. SV(P)/SVP then contains a polarizing p function (coefficients 0.1206750 and 1.0000000) as well and that basis set can be found under ``Ahlrichs pVDZ''. The Hartree--Fock, CCSD and CCSD(T) energies in a CAS of 24 electrons in 30 orbitals (freezing the lowest occupied orbitals) were compared using the Psi4 code. The Hartree--Fock was
-2085.57297 $E_\mathrm{h}$ in the SV basis and -2085.60285 $E_\mathrm{h}$ in the SV(P)/SVP basis. Our full active space SV CCSD(T) energy, -2086.39864 $E_\mathrm{h}$, agrees with Olivares-Amaya et al.\cite{Olivares-Amaya2015}. In this section, the correlation energies of other studies were calculated by subtracting the Hartree--Fock energy in a SV basis (no polarising p) off the total energy quoted in the various studies. The difference in correlation energies between the SV and the SV(P)/SVP basis sets with respect to the SV Hartree--Fock energy in this (24e, 30o) CAS was -0.03 and -0.05 for CCSD and CCSD(T) respectively. This difference is an order of magnitude larger than energy differences to those studies in this chromium investigation here. We therefore concluded that the basis set used was SV in Refs.\citenum{Sharma2012} and \citenum{Booth2014} as well.}.
\begin{figure}
	\includegraphics[width=8.5cm,keepaspectratio]{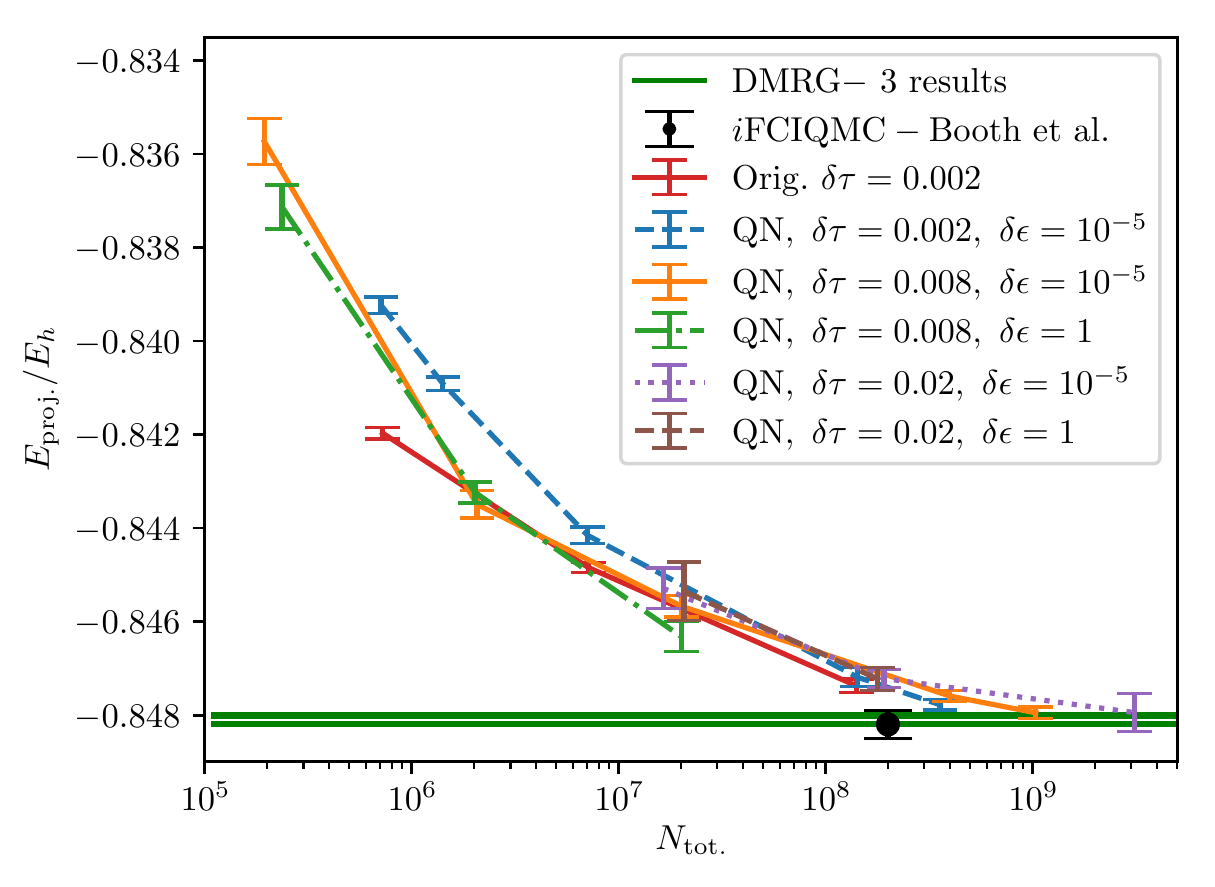}
        \caption{\small{Initiator curve of \ce{Cr2} in a (24 electrons, 30 orbitals) active space in SV basis\cite{Schafer1992} at a bond length of 1.5\AA. Three DMRG results\cite{Kurashige2009,Sharma2012,Olivares-Amaya2015} are shown with horizontal
            lines. An initiator curve point from Booth et al.\cite{Booth2014} is included. $\Delta_\mathrm{v} = 1$, $\rho=1$ here.}}
	\label{fig-Cr2-CAS}
\end{figure}
For the smaller CAS system, figure \ref{fig-Cr2-CAS} shows various initiator convergence curves, displaying
energy as a function of population size, for quasi-Newton and original propagation. The quasi-Newton
propagation was tested at $\delta\tau = 0.002, 0.008$ and $0.02$, whereas the original was only
stable or did not converge very slowly at $\delta\tau = 0.002$ out of these time steps (given the set initial population etc.). The range of the result by Booth et
al.\cite{Booth2014} is shown. Reblocking analysis was used to estimate errors on quoted energy values\cite{Flyvbjerg1989}. All initiator curves tend to this result and the threshold
$\delta\epsilon$ did not seem to have a noticeable effect.
\par
\begin{figure}
	\includegraphics[width=8.5cm,keepaspectratio]{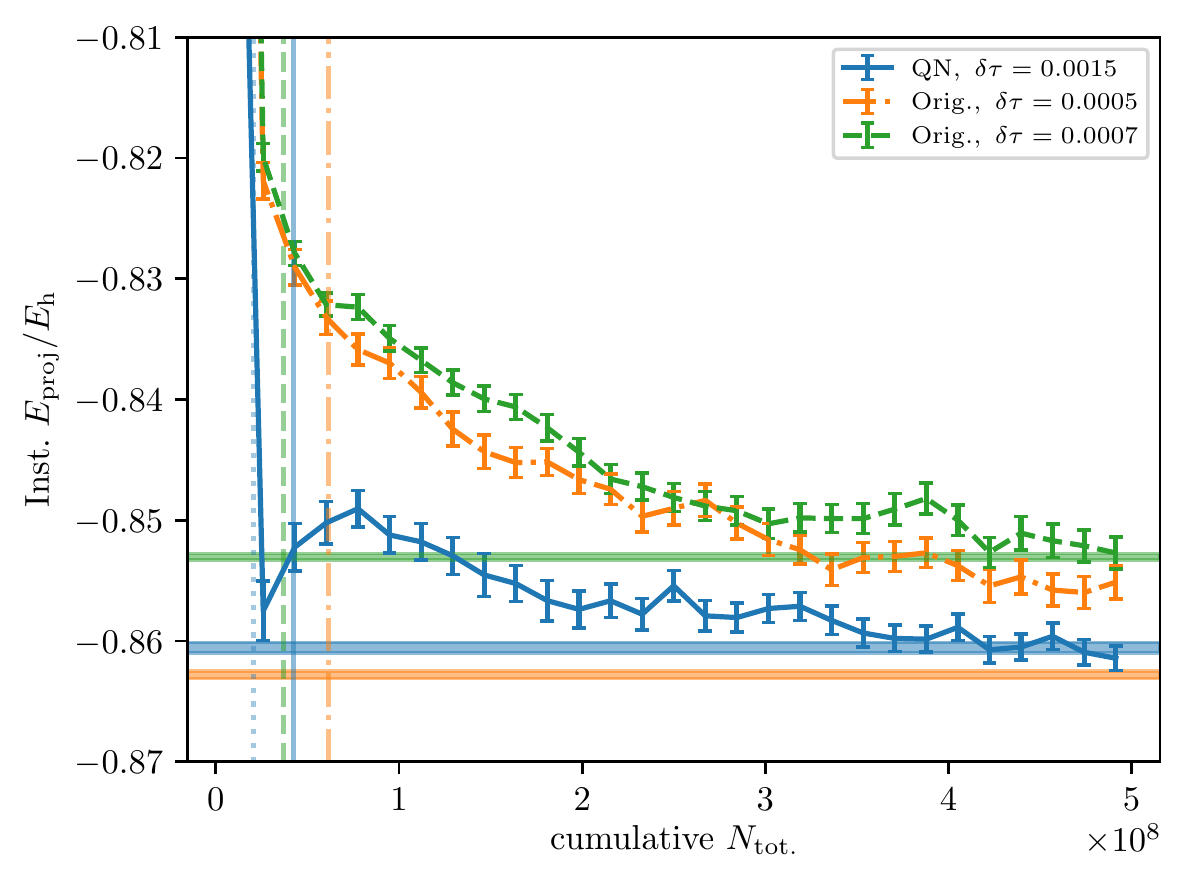}
	\caption{\small{Convergence of instantaneous projected energy in the all-electron chromium dimer in the SV basis\cite{Schafer1992} at a bond length of 1.5\AA\ of quasi-Newton (QN) and original propagation at a target population of 5$\times 10^5$ (population where shift starts varying) and initial population of 100 using initiator approximation. The QN results were run with $\rho = 0$ up to iteration 5000 ($\delta \tau = 0.0015$) and then set to $\rho = 1$, always using $\Delta_\mathrm{v} = 1$ and $\delta\epsilon=10^{-5}$. The instantaneous projected energy was binned with respect
		to the cumulative particle number and the bin means calculated. Each calculations was done at least 100 times
		in independent runs and the means of those bin means are shown with their 
		standard errors as error bars, placed at the number of cumulative
		particles that is at the middle of the bin. Only calculations where the inst. $E_\mathrm{proj.} < 0.1$ and $> -3.8 E_\mathrm{h}$ always were included. The horizontal lines (least negative $E_\mathrm{proj.}$ Orig. at $\delta\tau=0.0007$, then QN and most negative $E_\mathrm{proj.}$ is Orig. at $\delta\tau=0.0005$) indicate the mean $E_\mathrm{proj.}$ and its error found by taking the mean energy of all calculations of that type with at least 5$\times 10^5$ iterations. The left most vertical shows when $\rho$ was changed in the QN calculation and the others show when the shift was allowed to vary in the respective calculation. The vertical lines do not show error bars.}}
	\label{fig-Cr2-Full-conv}
\end{figure}

 Convergence of the full all-electron system with a Hilbert space size
of $10^{22}$ was then studied with initiator FCIQMC for a particular target population comparing quasi-Newton to original propagation (figure \ref{fig-Cr2-Full-conv}).
In figure \ref{fig-Cr2-Full-conv}, the convergence of original (at $\delta\tau=0.0007$) and quasi-Newton propagation defined as the point of overlap with the expected value is comparable. However, the quasi-Newton propagation is slightly faster convergent, even according to that definition, and the cost to get within $\pm 0.005 E_\mathrm{h}$, even if not stable, is significantly less costly than with the original propagation. 
\\An initiator quasi-Newton study with populations up to just above $10^{9}$ was done and a function of the form $a + bx^{-c}$
was fitted to the data set. The determined convergence value is -0.8717(3) $E_\mathrm{h}$ which agrees with DMRG\cite{Olivares-Amaya2015}, -0.871813 $E_\mathrm{h}$.
The maximum number of particles is of order $10^{9}$, a factor of $10^{13}$ reduction from the complete Hilbert space. As with the smaller CAS study, this shows that FCIQMC with quasi-Newton propagation gives reliable energies.

\section{Conclusion}
We have shown that the quasi-Newton propagation introduced here (applicable to both CCMC and FCIQMC) can accelerate the convergence of the (instantaneous) projected energy
compared to the original propagation, especially in large systems with a wider range in orbital energies. It scales more favourably ($\mathcal{O}(1)$ instead of $\mathcal{O}(N_\mathrm{el.})$) than the Jacobi propagation while having a comparable benefit. In conjunction with an excitation generator that does not scale with system size, such as the \textit{heat bath Power Pitzer ref.} excitation generator\cite{Neufeld2018a} in the case of CCMC, not adding extra scaling to the algorithm is important in large electronic systems. Using the quasi-Newton propagation, we quoted the first (initiator) FCIQMC result on the chromium dimer in the full SV basis set\cite{Schafer1992}.

\begin{acknowledgement}
The open-source HANDE code\cite{Spencer2019c} was used for stochastic and FCI calculations and Matplotlib\cite{Hunter2007} for figures.
We thank Dr. Nick Blunt and Mr. Bang Huynh for helpful conversations. Further information and the research data including (a link to) the deterministic propagation code will be available at \url{https://doi.org/10.17863/CAM.40122} after acceptance. Before then, please contact the authors for access.
V.A.N. would like to acknowledge the EPSRC Centre for Doctoral Training in Computational Methods for Materials Science for funding under grant number EP/L015552/1 and the Cambridge Philosophical Society for a studentship.
A.J.W.T. acknowledges the Royal Society for a University Research Fellowship under grants UF110161 and UF160398.
This work used the ARCHER UK National Supercomputing Service (\url{http://www.archer.ac.uk}) and the UK Research Data Facility (\url{http://www.archer.ac.uk/documentation/rdf-guide}) under ARCHER leadership grant e507.

\end{acknowledgement}

\providecommand{\latin}[1]{#1}
\providecommand*\mcitethebibliography{\thebibliography}
\csname @ifundefined\endcsname{endmcitethebibliography}
  {\let\endmcitethebibliography\endthebibliography}{}

\end{document}